\pdfoutput=1
\documentclass[a4paper, 10pt, USenglish, times, 1p]{elsarticle}%
\biboptions{sort}%
\usepackage[utf8]{inputenc}
\usepackage{babel}%
\usepackage{amsmath}%
\usepackage{amssymb}%
\usepackage{subcaption}%
\usepackage[separate-uncertainty=true, per-mode=symbol]{siunitx}%
\usepackage{tensor}%
\usepackage{hyperref}%
\usepackage[nameinlink,noabbrev]{cleveref}%
	\DeclareRobustCommand{\abbreviatedcleverreferences}{%
		\crefname{equation}{eq.}{eqs.}%
	}%
	\DeclareRobustCommand{\scref}[1]{{\abbreviatedcleverreferences\cref{#1}}}%
\renewcommand{\d}{\mathop{}\!\mathrm{d}}%
\renewcommand{\v}[1]{\protect \underline{#1}}%
\newcommand{\average}[1]{\langle#1\rangle}
\newcommand{\Parallel}{{\,/\!\!/\,}}%
\makeatletter%
\def\ps@pprintTitle{%
	\let\@oddhead\@empty%
	\let\@evenhead\@empty%
	\def\@oddfoot{\footnotesize© 2019. Licensed under a \href{http://creativecommons.org/licenses/by-nc-nd/4.0}{Creative Commons Attribution-NonCommercial-NoDerivatives 4.0 International License}.}%
	\let\@evenfoot\@oddfoot%
}
\makeatother%
\begin{document}%
\begin{frontmatter}%
	\title{\texorpdfstring{Fusion hindrance effects in\\laser-induced non-neutral plasmas}{Fusion hindrance effects in laser-induced non-neutral plasmas}}%
	\author[1,2,3,4]{\texorpdfstring{Salvatore Simone Perrotta\corref{cor1}}{Salvatore Simone Perrotta}}%
	\author[2,4]{Aldo Bonasera}%
	\cortext[cor1]{Corresponding author at: Dipartimento di Fisica e Astronomia, Università degli studi di Catania, via S.~Sofia 64, 95123 Catania, Italy.}%
	\address[1]{Scuola~Superiore~di~Catania, Università degli studi di~Catania, via Valdisavoia 9, 95123 Catania, Italy}
	\address[2]{Cyclotron Insitute, Texas A\&M University, 77843 College Station, TX, USA}
	\address[3]{Dipartimento di Fisica e Astronomia, Università degli studi di Catania, via S.~Sofia 64, 95123 Catania, Italy}
	\address[4]{Laboratori~Nazionali~del~Sud, Istituto Nazionale di Fisica Nucleare, via S.~Sofia 62, 95123 Catania, Italy}
	\begin{abstract}%
		Inertial confinement fusion hotspots and cluster Coulomb explosion plasmas may develop a positive net electric charge.
		The Coulomb barrier penetrability and the rate of nuclear fusion reactions at ultra-low energies ($\lesssim \SI{10}{\keV}$) are altered by such an environment. These effects are here studied via the screening potential approach.
		Approximate analytical results are developed by evaluating the average screening potential for some scenarios of interest.
		It is found that fusion is hindered for reactions between thermal fuel nuclei, while an enhancement is expected for secondary and “beam-target” reactions.
		Depending on the plasma conditions, the variations can be relevant even for relatively small net charges
		(several \si{\percent} difference or more in the fusion rate for an average net charge per nucleus of \num{e-5} proton charges).
	\end{abstract}%
\end{frontmatter}%
\section{Introduction}
	
	The field of nuclear fusion energy production is concerned with the search of conditions, attainable on the earth, that maximize the rate of suitable exothermic fusion reactions.
	To employ those as an energy source, a configuration with positive power balance (i.e.\ where the energy output is greater than the required input) must first be obtained.
	In particular, laser-induced inertial confinement fusion \cite{Pfalzner2006} and Coulomb explosion \cite{Lattuada2016,Zhang2017,Quevedo2018} facilities attempt to achieve the purpose by irradiating a target with laser beams, heating up a portion of fuel nuclei to temperatures of few \si{\keV}.
	In such conditions, the cross-section of nuclear reactions induced by charged particles
	depends exponentially on the collision energy, as the process is dominated by the tunneling through the Coulomb barrier \cite[sec.\ 3.2.1]{Iliadis2007}.
	Consequently, even relatively small energy differences may induce sensible variations in the cross-section.
	
	Most treatments for nuclear reactions in inertial confinement fusion systems
	assume that the hotspot is locally neutral (see e.g.\ \cite{Ramis2016,Hoffman2015}).
	This work explores the possibility that those plasmas may instead build up a positive net charge in their core.
	The study was also applied to Coulomb explosion systems, where non-neutral plasmas are instead certainly present, because they act as the acceleration mechanism.
	The effects of said charged mediums have been studied via the screening potential approach, i.e.\ by evaluating the energy transferred from the environment during reactants approach and available for the collision (\emph{screening potential}).
	The system was described as sphere with uniform and constant composition, net charge density and reactants temperature.
	Ions mean free path against significant deflections in the plasma was employed to define the distance covered by reactants under the environmental influence before colliding.
	The reaction rate modifications have been estimated by computing the average screening potential for the system, simplifying the problem and allowing for analytical solutions to be drawn.
	Refer to the thesis \cite{PerrottaTesiSSC} for further details on the model here presented.
	
	The paper is structured as follows.
	\Cref{sezscreeningpotentialapproach} outlines some useful notions regarding the screening effects on Coulomb penetrability in low energy reactions described via the screening potential approach,
	while \cref{sezIntroNonneutralplasmas} presents the state of the art on non-neutral plasmas for fusion energy production, with a focus on inertially confined systems.
	\Cref{sezModelDescription} describes the model developed for the proposed study, and
	\cref{sezModelApplication} shows its application to some scenarios of interest.
	In particular, the rate per particle pair and average cross-section alterations
	are explicitly computed for primary (“beam-beam”) reactions between a pair of identical reactants in \cref{sezUPrimaryReactions}.
	Finally, \cref{sezSummary} summarizes the main conclusions of this work.

\subsection{The screening potential approach}\label{sezscreeningpotentialapproach}
	
	Let $q_e$ be the electron charge modulus. Consider a fusion reaction induced by two nuclei of charge $Z_1 q_e$ and $Z_2 q_e$ and reduced mass $m$,
	taking place at a (non-relativistic) center-of-mass collision energy $E$ sensibly smaller than the maximum of their Coulomb barrier.
	Its total cross-sec\-tion $\sigma(E)$, dominated by the contribution of angular momentum $l=0$, can be written in terms of the \emph{astrophysical $S$-factor} $S(E)$ (see e.g. \cite[eq.~3.71]{Iliadis2007}):
	\begin{equation}\label{eqDefSFactor}
	\sigma(E) = S(E) \frac{1}{E} e^{- 2 \pi \eta(E)} \qquad , \qquad
	\eta(E) = \alpha_e Z_1 Z_2 \sqrt{\frac{m c^2}{2 E}}
	\end{equation}
	where $\alpha_e$ is the fine-structure coupling constant and $c$ the speed of light.
	The exponential term, representing the Coulomb barrier penetrability, accounts for the most prominent features of $\sigma(E)$ and causes it to vary sensibly even for relatively small differences in the collision energy.
	$S(E)$ is (at small $E$) a much more slowly varying function of $E$ than $\sigma$, especially for non-resonant reactions.
	
	Suppose that during the collision reactants are also interacting with the surrounding environment, which influences the Coulomb barrier penetration but has a negligible effect on the nuclear process.
	A widely studied example of this situation is found when reactant nuclei carry bound electrons, which screen the nuclear charge (the effect is treated in the sudden and adiabatic limits in \cite{Bracci1990}).
	Another well-known instance is that of neutral plasmas, where nuclei locally polarize other unbound ions and electrons, effectively screening their own charge (the classical treatment of the problem is given in \cite{Salpeter1954}).
	The \emph{effective collision energy} $E'$ is the energy value relevant for reaction dynamics, i.e.\ the reactants kinetic energy in their center-of-mass frame extrapolated to infinite distance treating the particles as isolated.
	As a result of the environmental interactions, the reactants effective collision energy changes as they approach: one may write $E'(r)$ as a function of reactants distance.
	This fact is usually accounted for in an effective manner, by treating the reaction as happening in vacuum and introducing a modification of the initial collision energy $E$ by an appropriate \emph{screening potential} $U$. Precisely, the \emph{screened} cross-section $\sigma_s$ (the one observed in the environment) is expressed in terms of the \emph{bare} cross-section $\sigma_b$ (the one for isolated nuclei) as:
	\begin{equation}\label{eqCrossSectionWithScreeningPotential}
	\sigma_s(E) = \sigma_b(E+U)
	\end{equation}
	$U$ represents the total kinetic energy acquired by reactants in their center-of-mass frame through environmental interaction during the collision, $U = E'(0) - E$, and can in general be a function of $E$ itself.
	The approach of \cref{eqCrossSectionWithScreeningPotential} is accurate when the energy transferred to reactants within the tunneling of their Coulomb barrier is negligible compared with $E+U$ (i.e.\ most energy is transferred at distances greater than the distance of classical closest approach).
	A sufficient condition is, being $U$ the total energy transfer:
	\begin{equation}\label{eqValidityConditionScreeningPot}
	\frac{|U|}{E+U} \ll 1
	\end{equation}
	
	Consider now a system of nuclei obeying an energy distribution $\Phi(E)$ (e.g.\ the fuel in a plasma).
	The fusion processes taking place herein are described by the \emph{reaction rate per particle pair} and the \emph{average cross-section}:
	\begin{equation}\label{eqDefineRatePerParticlePair}
	\average{\sigma v} = \int_{0}^{+\infty} \Phi(E) \sigma(E) \sqrt{\frac{2 E}{m}} \d E \qquad , \qquad
	\average{\sigma} = \int_{0}^{+\infty} \Phi(E) \sigma(E) \d E
	\end{equation}
	Assume that $\Phi(E)$ is a Maxwell-Boltzmann distribution (see e.g.\ \cite[eq.~3.7]{Iliadis2007}) with temperature $T$, and
	let $E_0 = [ \pi \eta(E) \sqrt{E} \, T ]^{2/3}$ (independent of $E$, see \scref{eqDefSFactor}; $E_0$ is the energy corresponding to the Gamow peak maximum).
	Further assume that the screening potential approach is applicable (basically, that $|U| \ll E_0$),
	that $U$ is independent of $E$,
	and that the astrophysical factor $S(E)$ can be approximated as constant near $E_0$ (the reaction is non-resonant).
	Then, the screened to bare ratios of the quantities in \cref{eqDefineRatePerParticlePair},
	i.e.\ the \emph{enhancement factors} $f$, can be calculated in the saddle point approximation (see e.g.\ \cite[sec.~2.2.1, 2.5.1]{Butler2007}),
	by inserting both \cref{eqDefSFactor,eqCrossSectionWithScreeningPotential}:
	\begin{equation}\label{eqScreenedRateSaddle}
	f_{\sigma v} = \frac{\average{\sigma_s v}}{\average{\sigma_b v}} \approx \left(1 - \frac{U}{E_0} \right) e^{U/T} \qquad , \qquad
	f_{\sigma} = \frac{\average{\sigma_s}}{\average{\sigma_b}} \approx \sqrt{1-\frac{U}{E_0}} e^{U/T}
	\end{equation}
	which in literature are normally reported neglecting the $U/E_0$ correction (see e.g.\ \cite[eq.~7]{Salpeter1954}).

\subsection{Non-neutral laser-induced plasmas for fusion energy production}\label{sezIntroNonneutralplasmas}

	A number of different techniques have been developed in the attempt to recreate, in laboratory, conditions somehow similar to those in stellar plasmas.
	The present work is particularly concerned with two laser-induced plasma fusion mechanisms, which share some experimental features: inertial confinement fusion and cluster Coulomb explosion.
	Their working principle is now briefly described.
	
	An \emph{inertial confinement fusion} (ICF) plasma (see e.g.\ the introductory book \cite{Pfalzner2006} or papers as \cite{Seguin2003,Hurricane2014,Csernai2015,Regan2016,Csernai2018,Zhang2018,Csernai2019,Gopalaswamy2019})
	is obtained by irradiating a target, either solid or enclosed in an external shell, containing fuel for the purported fusion reaction, with several laser beams.
	Electrons in the target surface receive the photons energy and subsequently accelerate also ions.
	The target outer layers are vaporized by the burst, and the explosion compresses and heats the inner layers.
	In the hotspot ignition concept, a central region (\emph{hotspot}) forms in the target, sensibly hotter and less dense than the rest of the compressed fuel.
	Using an appropriate configuration, enough fusion reactions may start in the hotspot
	during the \emph{confinement time} (i.e.\ before it disassembles itself driven by the high pressures formed),
	heating it up further and propagating the burn throughout the target.
	
	Inertial confinement fusion systems are often described by hydrodynamic numerical simulations.
	The plasma is in those modeled as two fluids in local thermodynamic equilibrium, representing ions and electrons, that evolve taking into account the interaction with lasers and nuclear reactions (see e.g.\ \cite{Ramis2016}). Several different models also exist (as \cite{Hoffman2015} or e.g.\ those referenced in \cite{Hurricane2014}).
	In all of them, it is assumed that the hotspot is locally neutral, i.e.\ the net charge density of ions and electrons is zero everywhere.
	The only (to the authors' knowledge) exception is made by a small-scale molecular dynamics simulation, reported in \cite{Bonasera2004Plasma}, where the neutrality condition is absent.
	The results of this work suggest that electrons, being much lighter than ions, could quickly reach the maximum compression and dilute again even before ions compress enough to start fusions, thus leaving a positively charged hotspot.
	
	It is instead observed in several works (see e.g.\ \cite[sec.\ VI.A.1]{Seguin2003} and references therein) that the target external layers can get positively charged, at least shortly after the laser pulse end.
	A large number of electrons are freed on the target surface (\emph{corona}) by laser interaction, and many of them rapidly leave the target, later followed by repelled ions.
	In experiments where \emph{bang time} (onset of fusion reactions) occurs before turning off the laser, fast ions emitted from the target consequently obtain a kinetic energy significantly greater than otherwise expected.
	The measured energy gain is proportional to the ion charge, and is usually smaller for bang times occurring later with respect to the laser pulse (see e.g.\ \cite[fig.\ 11]{Hicks2000}), suggesting that the target is gradually discharging.
	The target is normally assumed to be charged only far away from its center.
	Therefore, the electric field intensity would be progressively reduced only by expelling ions and by target expansion, possibly including electrons re-absorption (and perhaps by neutralizing currents from the target holder).
	Nevertheless, it seems reasonable that free electrons from internal layers may diffuse up to the corona, diluting the outer layers charge. It is even conceivable that a positive charge may persist in the hotspot for some time after the whole target already became globally neutral, if a sufficient amount of relatively slow electrons can be found lingering in the external regions.
	These electrons would be either previously emitted from the surface and then caught back, or formed with a mechanism as the one in \cite{Bonasera2004Plasma} mentioned above.
	Any effect due to a charged hotspot is anyway expected to fade away in experiments where the bang time is significantly delayed.

	In the \emph{Coulomb explosion} (CE) concept (see e.g.\ \cite{Lattuada2016,Zhang2017,Quevedo2018}), instead, the fuel is allowed through a nozzle to rapidly expand to average densities of $\sim \SI{e18}{atoms\per\centi\metre\cubed}$ \cite{Lattuada2016} (to be compared with ICF hotspot densities, e.g.\ $\sim \SI{e25}{atoms\per\centi\metre\cubed}$ in the measurements in \cite{Hurricane2014} at the NIF facility), forming molecular clusters.
	The radiation coming from a short-pulse laser removes electrons from clusters, whose ions thus get accelerated primarily by Coulomb repulsion with other members of the exploding cluster. Fusion reactions then take place between nuclei from different clusters.
	For comparison, recall that in an inertial confinement fusion plasma the hotspot often does not directly receive laser energy and is heated and compressed mainly by the external layers pressure.
	In the Coulomb explosion case, a charged environment is evidently present, but a model (known to the authors) that explicitly takes into account its effects on the fusion cross-sections, other than its role as ions acceleration mechanism, is nonetheless missing.

\section{Screening potential calculation scheme}\label{sezModelDescription}

	This section starts with a simple example, practically showing how the screening potential for a given configuration will be calculated.
	Then, it introduces the model employed in the present work to explore the effects that a given positively charged plasma may induce on the observed fusion cross-sections.
	The investigation on the charge formation mechanism is beyond the scope of this paper. To perform it would require, especially for ICF systems, a detailed hydrodynamic model, not relying on the charge neutrality hypothesis and capable of predicting the out-of-equilibrium system evolution over time as a function of the laser and target initial parameters.

\subsection{Prelude: point-like spectator charge}\label{sezModelPrelude}

	To expose some ideas behind this work, consider at first, instead of the physical scenario of interest, the following simplified example.
	Two identical particles with charge $Z q_e$ start at positions $\v r_+$, $\v r_-$ and collide in the proximity of a third fixed point-like charge $Z_3 q_e$ in the origin (or, equivalently, a spherically-symmetric charge distribution that is never penetrated by the other ions).
	Reactants center-of-mass is at rest in $\v r_f$ in $Z_3$ rest frame.
	The electrostatic potential energy induced by $Z_3$ on a single reactant in a position $\v r$ is
	%
	$V(\v r) = + k_e q_e^2 Z Z_3 / r$,
	where $k_e q_e^2 \approx \SI{1.44}{\MeV \femto\metre}$.
	Thus, reactants will change their collision energy while approaching, because of the interaction with $Z_3$.
	The relative energy difference is normally quite small (yet relevant for \cref{eqDefSFactor}),
	consequently, ions trajectories are assumed for simplicity to be the same as they would for $Z_3 = 0$.
	
\begin{figure}[tbp]%
	\begin{subfigure}[b]{.5\linewidth}%
		\centering
		\includegraphics[keepaspectratio = true]{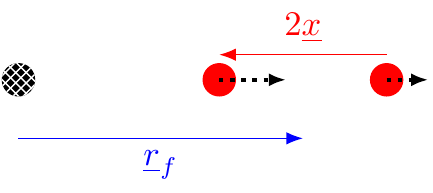}%
		\caption{\label{fig_SimpleExamplesParallel}$\v r_f \Parallel \v x$}
	\end{subfigure}%
	\begin{subfigure}[b]{.5\linewidth}%
		\centering
		\includegraphics[keepaspectratio = true]{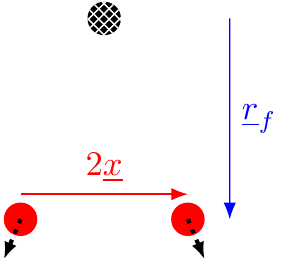}%
		\caption{\label{fig_SimpleExamplesPerp}$\v r_f \perp \v x$}
	\end{subfigure}%
	\caption{\label{fig_SimpleExamples}%
		Schematic representation of the examples discussed in \cref{sezModelPrelude}. In both panels, the crosshatched black point is the fixed point-like spectator charge, and the full red points signal
		reactants initial position. Dotted black arrows roughly represent the force exerted by the spectator on reactants (particle velocities are not drawn). $\v r_f$ is the reactants center-of-mass position with respect to the spectator and $2 \v x$ is the initial reactants distance.}
	\end{figure}

	Let $2 \v x = \v r_- - \v r_+$ be the initial displacement between the reactants pair, and consider in particular the two configurations depicted in \cref{fig_SimpleExamples},
	i.e.\ with $\v r_f$ either parallel or perpendicular to $\v x$.
	In the former case it is $\left|\v r_\pm\right| = r_f \pm x$, thus, while reactants move from $\v r_\pm$ to $\v r_f$, they gain in $Z_3$ rest frame a kinetic energy (due only to the environmental interaction):
	\begin{equation}
	\Delta E_{\pm} = - \Delta V_{\pm} = k_e q_e^2 Z Z_3 \left(\frac{1}{r_f \pm x} - \frac{1}{r_f}\right) = \mp k_e q_e^2 Z Z_3 \frac{x}{(r_f \pm x) r_f}
	\end{equation}
	For $r_f > x$, one reactant gains kinetic energy and the other loses it (since one moves away from $Z_3$ and the other comes closer), but the total gain:
	\begin{equation}
	\Delta E = \Delta E_+ + \Delta E_- = k_e q_e^2 Z Z_3 \frac{2x^2}{r_f (r_f^2 - x^2)}
	\end{equation}
	has the sign of $r_f - x$, i.e.\ it is positive unless $Z_3$ actually lies between reactants initial positions.
	
	Since the reactants center-of-mass is initially at rest, its environmental energy gain is negligible in the limit of $\left|\Delta E_{\pm}\right| / E_\pm \to 0$ (being $E_\pm$ the initial reactants energies, see \cref{eqPotScreenrestCMSecondOrder}).
	Therefore, $\Delta E$ here approximately coincides with the screening potential $U$,
	and the configuration shown in \cref{fig_SimpleExamplesParallel} enhances fusion for any $r_f > x$.
	
	Instead, in the case of \cref{fig_SimpleExamplesPerp} it is $\left|\v r_\pm\right| = \sqrt{r_f^2 + x^2} > r_f$, hence also the energy gains are equal,
	\begin{equation}
	\Delta E_\pm = k_e q_e^2 Z Z_3 \left(\frac{1}{\sqrt{r_f^2 + x^2}} - \frac{1}{r_f}\right)
	\end{equation}
	and always negative: fusion is hindered in this configuration.
	In both cases, the screened cross-section for a given initial collision energy $E$ is then given by \cref{eqCrossSectionWithScreeningPotential}, and can be calculated explicitly using e.g.\ \cref{eqDefSFactor}.
	
	Incidentally, since the electrostatic potential is additive, it can be shown that the energy gains here calculated (and thus the screening potential) correspond to the one given in \cref{sezReactantscenterofmassatrest}, once they are summed over all possible positions for $Z_3$ inside a sphere. Such a sum effectively represents an uniform charge distribution, or, equivalently, the average effect of a large number of point charges (Coulomb explosion clusters) randomly distributed in the sphere.

\subsection{Inertial confinement fusion systems}\label{sezModelICF}
	
	The proposed model is here described making explicit reference to inertial confinement fusion plasmas. The few modifications required to apply it to Coulomb explosion systems are illustrated in \cref{sezModelCE}.
	
	The hotspot is described as a sphere of constant and uniform mass, net charge, and isotopic abundance distribution (any spherically symmetric charge profile outside the sphere is irrelevant).
	Reactants (always treated as non-relativistic) follow a Maxwell-Boltzmann energy distribution, with a temperature that can be set equal to the one experimentally extracted from emitted neutrons energy spectrum (since they are not influenced by the system charge).
	This configuration is employed to estimate a screening potential for the system, and in particular to deduce how this and the consequent rate and average cross-section enhancement factors vary with plasma parameters.
	
	Note that the environment charge density could be perturbed locally by reactants polarization, as it happens in a neutral plasma \cite{Salpeter1954}.
	It is also expected that the temporary presence of a net charge will somewhat reduce the system confinement time, affecting the measured fusion yields. Those effects are at present ignored.
	Furthermore, if reactants were not fully ionized, their atomic screening effects would have to be taken into account separately (adding the different screening potentials is usually accurate enough), as this work is clearly not concerned with those.
	
	Let $R$ be the sphere radius, $Z_{\text{tot}} q_{e}$ its total net electric charge ($q_{e}$ is the proton charge), and $\tilde\rho_{s} = Z_{\text{tot}}/R^{3}$, so that $q_{e} \tilde\rho_{s} / (\frac{4}{3} \pi)$ is the sphere net charge density.
	The electrostatic potential energy, denoted $V$, of a point charge $q_e$ inside the sphere as a function of the displacement $\v r$ from its center is (setting $V(0)=0$):
	\begin{equation}\label{eqElectrostaticPotentialUniformSphere}
	V(r) = - \frac{1}{2} k_e q_e^2 \tilde\rho_s r^2
	\end{equation}
	Consider then two reactants, arbitrarily marked in the following as “$+$” and “$-$”,
	starting at positions $\v r_+$ and $\v r_-$ and ending as a fused nucleus in $\v r_f$.
	During the collision, these particles will gain, due to the interaction with the charged environment, in the sphere rest frame, an energy:
	\begin{equation}\label{eqEnergyGainFixedPotential}
	\Delta E = \Delta E_+ + \Delta E_- = Z_+ \left[ V(r_+) - V(r_f) \right] + Z_- \left[ V(r_-) - V(r_f) \right]
	\end{equation}
	where $\Delta E_\pm$ are the individual contributions due to the two reactants, and $Z_\pm$ are the particles charge numbers.
	Part of this energy is spent to accelerate the reactants center of mass: let $\Delta E_{\text{CM}}$ be its energy gain in the sphere rest frame. The screening potential is thus:
	\begin{equation}\label{eqUExpression}
	U = \Delta E - \Delta E_{\text{CM}}
	\end{equation}
	
	Even though the reactants energies change, their velocities vary much less during most
	of their approach. Hence, they are approximated as constants (“free-motion assumption”) only for the purpose of computing the particles trajectory in the plasma (as done in \cite{Bracci1990} to evaluate the atomic electrons screening), greatly simplifying the problem.

\subsubsection{Mean free path and initial reactants distance}\label{sezInitialReactantsDistance}
	
	Reactants initial positions are related to their time of last “equilibration” with the medium via scattering with other particles, i.e.\ the last time when their kinetic energy obeyed the global energy distribution.
	In a neutral plasma, the total potential felt by a pair of reactants rapidly falls to zero as their distance increase, thus the concept of initial position is unneeded ($\Delta E$ is the same for any big enough $|\v r_+ - \v r_-|$).
	Here, the total potential and the environmental interaction are non-negligible at all distances, so it is necessary to define to what configuration the initial collision energy $E$ (in turn connected to the system temperature $T$) corresponds.
	For instance, it would be incorrect to set the initial positions at infinity, implying that ions come from outside the sphere with energy distributed according to the hotspot temperature, and then collide within it.
	The choice adopted in the present work is to link the total distance covered, on average, by a reactant $Z_\pm$ from initial to final position, $\average{\tilde d_\pm} = \average{|\v r_f - \v r_{\pm}|}$, to an appropriate mean free path for the system, $\lambda_\pm$.
	Each configuration will be given as weight the probability $\exp(- \average{\tilde d_\pm} / \lambda_\pm)$ for $Z_\pm$ to cover said distance without significant deflections.
	
	Let $m_+$ and $m_-$ be the reactants masses, $M$ their sum, $m = m_+ m_- / M$ the reduced mass,
	$E$ their initial collision energy in the center-of-mass frame,
	$\v P$ their center-of-mass momentum in the charged sphere rest frame,
	$E_{\text{CM}} = P^2/(2M)$, the ratio $\gamma = E_{\text{CM}}/{E}$,
	the initial reactants distance $2 \v x = \v r_- - \v r_+$, and $\theta_{Px}$ the angle between $\v x$ and $\v P$.
	An interval $t_c = 2 x / \sqrt{2 E / m}$ approximately elapses between the initial and fusion time
	(for a $\nuclide{d}+\nuclide{d}$ system at $E = \SI{1}{\keV}$, the relative correction taking into account the reactants mutual repulsion is of \SI{10}{\percent} or less for $x > \SI{10}{\pico\metre}$).
	Thus, the distance covered in the sphere rest frame by the reactants center-of-mass during the collision is:
	\begin{equation}\label{eqCenterOfMassCoveredLenght}
	d_{\text{CM}} \approx t_c \frac{P}{M} = \sqrt{\frac{4 m}{M} \gamma} \, x
	\end{equation}
	The distance $\tilde d_{\pm}$ covered by a reactant during the collision is then:
	\begin{equation}\label{eqReactantCoveredLenght}
	\tilde d_{\pm} = \sqrt{ d_{\text{CM}}^2 + \left(\frac{2 m_{\mp}}{M} x\right)^2 \pm 2 d_{\text{CM}} \frac{2 m_{\mp}}{M} x \cos\theta_{Px} }
	\end{equation}
	and its average over all $\v x$ and $\v P$ directions is:
	\begin{equation}\label{eqAngleAverageReactantCoveredLenght}
	\average{\tilde d_\pm} = \left\lbrace \begin{aligned}
	& \left( \frac{1}{3} \frac{m_\mp}{m_\pm} \frac{1}{\gamma} + 1 \right) d_{\text{CM}}
	&\text{if } & \gamma \geq \frac{m_\mp}{m_\pm} \\
	& \left( \frac{1}{3} \frac{m_\pm}{M} \gamma + \frac{m_\mp}{M} \right) 2 x
	&\text{if } & \gamma \leq \frac{m_\mp}{m_\pm}
	\end{aligned} \right.
	\end{equation}
	Note that $\average{\tilde d_\pm} / x$ does not depend on $x$.
	
	Regarding the mean free path values, for inertial confinement fusion systems the only possibility in the present work is
	referring to the case of a neutral plasma, for which extensive literature exists. The mean length covered by $Z_a$ before being deflected by \SI{1}{\radian} (calculated in the center-of-mass frame between $Z_a$ and the scatterers) as a result of several small scatterings on the medium is usually parametrized as (see e.g.\ \cite[eq.\ 9.38]{Eliezer2002}):
	\begin{equation}\label{eqMeanFreePathICF}
	\frac{1}{\lambda_a} = 2 \pi \left(\frac{k_e q_e^2 Z_a}{T}\right)^2 \sum_j Z_j^2 n_j \ln\Lambda_{aj}
	\end{equation}
	where $T$ is the system temperature, $n_j$ is species $j$ number density, $Z_j$ is the charge number, the sum is performed over all species in the medium (including electrons), and $\ln\Lambda_{aj}$ is the \emph{Coulomb logarithm} \cite[eq.\ 9.36]{Eliezer2002}.
	The exact value of $\ln\Lambda_{aj}$ depends on the ion and plasma proprieties, but, thanks to the smoothening operated by the logarithm, it is possible to approximate it with a constant. Within the range of plasma parameters (temperature, mass and net charge density) of interest in the present work, $\ln\Lambda = 10$ will be always adopted for ICF systems.
	
	This reasoning breaks down when $\lambda$ is comparable or greater than the hotspot size (e.g.\ about \SI{20}{\micro\metre} in \cite{Regan2016}).
	For one thing, the mean free path outside the hotspot has a different value (smaller, since in ICF plasmas the external regions are colder and denser).
	Also, reactants energy loss outside the hotspot can play an important role in hindering fusion.
	For another, the net charge distribution outside the hotspot can be different from the one inside, so the screening potential expressions calculated in the following sections are not accurate for high $\tilde d$, which become more probable for high $\lambda$.

\subsubsection{Average screening potential}
	
	For any given configuration of the system under study (plasma parameters, $\v r_+$, $\v r_-$ and $\v r_f$, particles energy\dots),
	it is possible to evaluate
	the screening potential $U$, via \cref{eqUExpression}, and the corresponding screened cross-section $\sigma_s$, via \cref{eqDefSFactor,eqCrossSectionWithScreeningPotential}.
	The observed screened rate per particle pair and screened average cross-section could then be found by properly averaging $\sigma_s$ on all possible configurations for the system, including the average on $\tilde d_\pm$ (defined in \cref{eqReactantCoveredLenght}) and on the reactants energy distribution.
	An effective screening potential $U_{\text{eff}}$ may be also extracted from the average $\sigma_s$ at fixed collision energy, but normally $U_{\text{eff}}$ depends on $E$ even if the $U$ for each possible configuration does not.
	This implies that $U_{\text{eff}}$ cannot be inserted in \cref{eqScreenedRateSaddle}.
	
	In practice, such a procedure is consequently rather complicated, and could be performed only numerically in most cases.
	In the present work, a simpler treatment is given:
	the average screening potential is computed (instead of averaging directly the cross-section), using $\average{\tilde d_\pm}$ in \cref{eqAngleAverageReactantCoveredLenght} for the weighting on the initial reactants distance (instead of $\tilde d_{\pm}$),
	\begin{equation}\label{eqExplicitxAverage}
	\average{U}_{|\lambda_+ ,\, \lambda_-}
	= \frac{
		\int_0^{+\infty} U \exp\left(- \average{\tilde d_+} / \lambda_+ - \average{\tilde d_-} / \lambda_- \right) \d x
	}{
		\int_0^{+\infty} \exp\left(- \average{\tilde d_+} / \lambda_+ - \average{\tilde d_-} / \lambda_- \right) \d x
	}
	\end{equation}
	If necessary, the resulting $\average{U}$ is further averaged on the reactants energy distributions, to obtain an energy-independent screening potential, which can be inserted in \cref{eqScreenedRateSaddle} to find the enhancement factors.

\subsection{Coulomb explosion systems}\label{sezModelCE}

	For the purposes of this work, there are three main differences between inertial confinement fusion and Coulomb explosion environments.
	First of all, the typical values of the plasma parameters, especially the density, differ in the two cases.
	
	Second, while in an ICF plasma all particles in the hotspot are heated up in the compression and thus obey a similar energy distribution, in a CE system several atoms are not ionized by the laser and do not explode.
	It is still possible to approximately define a temperature $T$ for the accelerated ions in the CE plasma (which is not the actual plasma temperature), but a population of cold fuel nuclei is also present in the environment.
	
	Most importantly, the particles spatial distribution is very different. In the ICF case the medium is well described by a continuous system with uniform density, but a CE plasma is made of relatively dense clusters separated by vast, nearly empty spaces.
	For this reason, the expression in \cref{eqMeanFreePathICF} for the ions mean free path in the plasma is certainly unsuitable for a CE system: the estimate developed in \cite{Zhang2017} will be adopted in its place.
	Assuming that a particle will travel freely until it penetrates a cluster, and then certainly be strongly deflected, $\lambda$ is identified with $1/(\sigma_c \rho_c)$, where $\rho_c$ is the density of clusters in the plasma and $\sigma_c$ is the (average) geometrical cross-section of a cluster.
	The density of clusters is $\rho_c = n_i / N_{ic}$, where $n_i$ is the total number density of nuclei and $N_{ic}$ the average number of nuclei per cluster, here set to \num{e3} (see \cite[figs.\ 2, 3]{Zhang2017}).
	The cross section is written as $\sigma_c = \pi R_c^2$, with $R_c = r_s N_{ic}^{1/3}$ the radius of a cluster (the impinging particles dimensions, considerably smaller, are ignored), and $r_s$ a typical length scale, here set to \SI{0.17}{\nano\metre} as in \cite{Zhang2017}.
	Hence:
	\begin{equation}\label{eqMeanFreePathCE}
	\lambda_{\text{CE}} = \frac{N_{ic}^{1/3}}{\pi r_s^2 n_i}
	\end{equation}
	This expression is independent of the incident ion properties and the plasma chemical composition, net charge and temperature, but has the same scaling as \cref{eqMeanFreePathICF} with the plasma density of nuclei.
	\Cref{eqMeanFreePathCE}, as \scref{eqMeanFreePathICF}, is not appropriate for $\lambda$ comparable or greater than the environment size, but note that Coulomb explosion systems are usually much bigger than ICF hotspots (up to several \si{\milli\metre} in radius \cite{Zhang2017,Quevedo2018}), thus the limitation is here less concerning.
	
	The net charge distribution, instead, will be here for simplicity assumed to be uniform and continuous, as in \cref{sezModelICF}, since only its large-scale features are taken into account in this work, and only the average screening potentials are calculated. It is expected that an accurate evaluation of the average enhancement factors for a more realistic distribution, e.g.\ a discrete set of randomly placed point-charges, would yield slightly higher cross-sections.
	In fact, greater fluctuations on $U$, for fixed $\average{U}$, favor in general greater enhancements, due to the exponential term in \cref{eqDefSFactor} (see \cite[sec.\ 1.2.A]{PerrottaTesiSSC} for further details).

\section{Enhancement factors estimations}\label{sezModelApplication}
	
	In this section, the model presented in \cref{sezModelDescription} is applied to three different reactant configurations.
	Those in \cref{sezUSecondaryReactions,sezUPrimaryReactions} directly correspond to physical scenarios of interest.
	The one in \cref{sezReactantscenterofmassatrest} (which is the extension to the example in \cref{sezModelPrelude} for the spectator charge distribution here of interest), on the other hand,
	has mostly a conceptual importance, due to its simplicity and the reduced number of approximations required to obtain the result.

\subsection{Secondary reactions}\label{sezUSecondaryReactions}

	Suppose one particle, say $Z_+$, is initially at rest with respect to the charged sphere. This is an appropriate limit to represent \emph{secondary reactions}, i.e.\ fusions between a fuel nucleus at thermal energy (or less) and a fast product of another previous reaction or scattering.
	In Coulomb explosions, a similar configuration also occurs for \emph{beam-target} reactions, i.e.\ between a ‘thermal' ion accelerated by a cluster and a ‘cold' atom.
	
	By the free-motion assumption, the fused nucleus position $\v r_f$ is approximately the initial $Z_+$ position, $\v r_+$ (notation as in \cref{sezModelDescription}).
	Given that both $\v r_+$ and $\v r_-$ lie inside the charged sphere, the energy gains in \cref{eqEnergyGainFixedPotential} are then $\Delta E_+ = 0$ (as $Z_+$ approximately does not move) and:
	\begin{equation}
	\Delta E_- = - 2 k_e q_e^2 Z_- \tilde\rho_s x r_0 \cos\theta_{x0}
	\end{equation}
	where $\v r_0 = (\v r_- + \v r_+)/2$ is the reactants initial middle point (with respect to the sphere center)
	and $\theta_{x0}$ is the angle between $\v r_0$ and $2 \v x = \v r_- - \v r_+$.
	The energy gained by the center-of-mass is $\Delta E_{\text{CM}} = \frac{m_-}{M} \Delta E_-$, thus the screening potential is:
	\begin{equation}\label{eqScreeningPotReactantRest}
	U = - \frac{2 m_+}{M} k_e q_e^2 Z_- \tilde\rho_s x r_0 \cos\theta_{x0}
	\end{equation}
	It is then necessary to average the result on all possible reactant configurations.
	The average of \cref{eqScreeningPotReactantRest} on $\cos\theta_{x0}$ (the $\v x$ directions) is simply:
	\begin{equation}\label{eqAverageScreeningPotReactantRest}
	\average{U}_{|r_0} = 0
	\end{equation}
	i.e.\ the approximate calculation predicts no effects. This means that a moderately enhanced fusion rate can be expected from an exact computation of the average enhancement factors, for the same reason mentioned at the end of \cref{sezModelCE}.
	
	Note that the chosen parametrization for $U$ has a relevance in determining the reactants spatial distribution for the average.
	For instance, \cref{eqAverageScreeningPotReactantRest} is an average, at fixed $x$ and $\v r_0$, on different $\v x$ directions, followed by an (here trivial) average on $\v r_0$, which produces an equal distribution for $\v r_+$ and $\v r_-$ positions.
	Taking instead an average at fixed $\v r_+$ and different $\v x$ directions, followed by an average on $\v r_+$ in the sphere, would imply that $\v r_+$ is (on average) found nearer to the sphere center than $\v r_-$, resulting in a non-zero $\average{U}$.

\subsection{Reactants center-of-mass at rest}\label{sezReactantscenterofmassatrest}

	Assume now that the reactants initial total momentum is zero in the charged sphere rest frame, i.e.\ it is $E_{\text{CM}} = 0$ (notation as in \cref{sezInitialReactantsDistance,sezUSecondaryReactions}).
	The center-of-mass is then (by the free-motion assumption) at rest at $\v r_f = \v r_0 + \frac{m_2 - m_1}{M} \v x$, and the reactants energy gains are:
	\begin{equation}\label{eqEnergyGaincenterofmassatrest}
	\Delta E_{\pm} = - \frac{1}{2} k_e q_e^2 Z_{\pm} \tilde\rho_s \left( \frac{4 m}{M} x^2 \mp 2 r_0 x \frac{2 m_\mp}{M} \cos\theta_{x0} \right)
	\end{equation}
	The average energy gain between any pair of configurations with opposite $\v x$, at given $x$ and $r_0$, is:
	\begin{equation}
	\average{\Delta E_\pm}_{|r_0} = - \frac{1}{2} k_e q_e^2 \tilde\rho_s Z_{\pm} \frac{4 m}{M} x^2
	\end{equation}
	independent of $r_0$ and $\theta_{x0}$.
	After a trivial average over $\v r_0$, this corresponds to the average over all $\v r_+$, $\v r_-$ positions, at fixed $x$, extracted with equal distribution.
	
	The average total energy gain is then clearly $\average{\Delta E}_{|r_0} = \average{\Delta E_+}_{|r_0} + \average{\Delta E_-}_{|r_0}$.
	Note that if $m_+/m_- = Z_+/Z_-$ (e.g.\ for identical reactants), then $\Delta E_+ + \Delta E_- = \average{\Delta E}_{|r_0}$ always (not on average):
	fusion is then (for fixed $x$) hindered approximately in the same way regardless of the reactants initial configuration (as long as they both lie inside the charged sphere).
	
	$\Delta E_{\text{CM}}$ is instead just the final center-of-mass energy. The center-of-mass momentum modulus at fusion time is in this case:
	\begin{equation}
	P' = \left| \sqrt{2 m_+ (E_+ + \Delta E_+)} - \sqrt{2 m_- (E_- + \Delta E_-)} \right|
	\end{equation}
	where $E_{\pm}$ are the ions initial kinetic energies in the sphere rest frame, extrapolated to infinite distance treating the reactants pair as isolated (as is done for the collision energy $E$, so that, in general, $E_+ + E_- = E + E_{\text{CM}}$). Thus:
	\begin{equation}
	\Delta E_{\text{CM}} = \frac{m_+}{M} \Delta E_+ + \frac{m_-}{M} \Delta E_- + \sqrt{\frac{4 m}{M} E_+ E_-} \left[ 1 - \sqrt{ \left(1 + \frac{\Delta E_+}{E_+} \right) \left(1 + \frac{\Delta E_-}{E_-} \right) } \right]
	\end{equation}
	If $\left|\Delta E_{\pm}\right| / E_\pm \ll 1$, a condition closely related to \cref{eqValidityConditionScreeningPot}, the square root may be expanded to second order, giving:
	\begin{equation}\label{eqPotScreenrestCMSecondOrder}
	U \approx \Delta E - \frac{1}{4 E} \left( \sqrt{\frac{m_+}{m_-}} \Delta E_+ - \sqrt{\frac{m_-}{m_+}} \Delta E_- \right)^2
	\end{equation}
	For $E$ high enough the second-order correction (caused by the center-of-mass acceleration) may be neglected, as done in the following, obtaining $U \approx \Delta E$.
	
	Further assume for simplicity that the mean free path $\lambda$ is the same for both reactants (which is always true using \cref{eqMeanFreePathCE}, and holds if and only if $Z_+ = Z_-$ using \cref{eqMeanFreePathICF}).
	Then, the average of $\average{U}_{|r_0}$ over $x$, carried out as in \cref{eqExplicitxAverage}, is:
	\begin{equation}\label{eqAverageScreenPotRestCM}
	\average{U}_{|r_0, \, \lambda, \, E_{\text{CM}} = 0} \approx - k_e q_e^2 \tilde\rho_s \frac{Z_+ + Z_-}{2} \frac{4 m}{M} \frac{\lambda^2}{2}
	\end{equation}
	independent of $E$.
	The corresponding enhancement factors are not shown here for brevity, since the screening potential found in \cref{eqScreenPotArbitraryMotionEnergyAveraged} for primary reactions differs from \cref{eqAverageScreenPotRestCM} only by a constant, thus all results are analogous.

\subsection{Primary reactions between identical reactants}\label{sezUPrimaryReactions}

	At last, consider a pair of reactants with arbitrary $E > 0$ and $\gamma = E_{\text{CM}}/{E}$ (notation as before).
	If the results are averaged distributing the energies on a Maxwell-Boltzmann distribution, they describe \emph{primary reactions} (\emph{beam-beam reactions} in Coulomb explosion systems), i.e.\
	between pairs of fuel nuclei at thermal energy.
	
	Said $\v r_i$ the initial position of the reactants center-of-mass (with respect to the sphere center), each energy gain in \cref{eqEnergyGainFixedPotential} can be written as:
	\begin{equation}
	\Delta E_{\pm} = Z_{\pm} \left[ V\left(\v r_{\pm}\right) - V\left(\v r_i\right) \right] + Z_{\pm} \left[ V\left(\v r_i\right) - V\left(\v r_f\right) \right]
	\end{equation}
	The first term is equal to \cref{eqEnergyGaincenterofmassatrest}, while the second one is, apart from the $Z_{\pm}$ factor, equal for both reactants.
	Using \cref{eqCenterOfMassCoveredLenght} for $d_{\text{CM}} = \left| \v r_f - \v r_i \right|$, it is:
	\begin{multline}\label{eqPrimaryReactionsReactantsEnergyGain}
	\Delta E_{\pm}
	= k_e q_e^2 Z_{\pm} \tilde\rho_s \left[ \pm x r_0 \frac{2 m_\mp}{M} \cos\theta_{x0} + \right. \\
	\left. + x \sqrt{\frac{4 m}{M} \gamma} \left( r_0 \cos\theta_{P0} + \frac{m_- - m_+}{M} x \cos\theta_{Px} \right) + \left( \gamma - 1 \right) \frac{2 m}{M} x^2 \right]
	\end{multline}
	where $\theta_{P0}$ is the angle between $\v r_0$ and the initial center-of-mass momentum $\v P$.
	
	The average gain is here computed as follows.
	First, $\Delta E_{\pm}$ is averaged over two configurations, at fixed $\v r_0$, with opposite $\v P$ and $\v x$ (thus with fixed $\theta_{Px}$ and opposite $\cos\theta_{P0}$ and $\cos\theta_{x0}$). Then, the result is summed over two pairs of configurations, at fixed $\v x$ and $\v r_0$, with opposite $\v P$: this second average is not necessary if $m_1 = m_2$.
	\begin{equation}
	\average{\Delta E_{\pm}}_{|r_0} = \frac{1}{2} k_e q_e^2 Z_{\pm} \tilde\rho_s \left( \frac{E_{\text{CM}}}{E} - 1 \right) \frac{4 m}{M} x^2
	\end{equation}
	Therefore, $\v r_1$ and $\v r_2$ are extracted (for every fixed $x$) from the same distribution,
	while $\v r_f$ is always extracted in a larger region than $\v r_0$ (i.e.\ the reactants middle point must start inside the sphere but may end up outside).
	This is in principle more appropriate, for the problem at hand, with respect to extracting $\v r_0$ and $\v r_f$ in the same region (so that, due to $x>0$, reactants must collide in a region smaller than the starting one).
	However, remind that the results found in this paper are not accurate if reactants step outside the charged sphere: a proper treatment of this possibility is expected to favor lower enhancement factors in inertial confinement fusion scenarios.
	
	Let $\v p_{\pm}$ be the vectors directed as $Z_\pm$ velocity and with norm $p_\pm^2 = 2 m_\pm E_\pm$
	(i.e.\ the “collisional” initial reactants momenta in the sphere rest frame).
	Similarly, let $\v p'_\pm$ be the vectors parallel to $\v p_{\pm}$ (by the free-motion assumption) and with modulus $p'_\pm = p_\pm \sqrt{1 + \Delta E_\pm / E_\pm}$ (i.e.\ the final momenta).
	The energy gained by the reactants center-of-mass, in the sphere rest frame, during the collision is such that:
	\begin{multline}\label{eqPrimaryReactionsCoMEnergyGainExact}
	M \Delta E_{\text{CM}}
	= \frac{1}{2} \left( \v p'_+ + \v p'_- \right)^2 - \frac{1}{2} \left( \v p_+ + \v p_- \right)^2 = \\
	= m_+ \Delta E_+ + m_- \Delta E_- + \v p_+ \cdot \v p_- \left[ \sqrt{ \left(1 + \frac{\Delta E_+}{E_+} \right) \left(1 + \frac{\Delta E_-}{E_-} \right) } - 1 \right]
	\end{multline}
	Provided that $\left|\Delta E_{\pm}\right| / E_\pm \ll 1$, the square root in \cref{eqPrimaryReactionsCoMEnergyGainExact} may be expanded to first order.
	Since $\v p_+ \cdot \v p_- = M E_{\text{CM}} - m_+ E_+ - m_- E_-$, it is then:
	\begin{equation}\label{eqPrimaryReactionsCoMEnergyGain}
	\Delta E_{\text{CM}} \approx \frac{E_{\text{CM}}}{2} \left( \frac{\Delta E_+}{E_+} + \frac{\Delta E_-}{E_-} \right) + \frac{E_+ m_+ - m_- E_-}{2 M} \left( \frac{\Delta E_+}{E_+} - \frac{\Delta E_-}{E_-} \right)
	\end{equation}
	where:
	\begin{equation}
	\v p_{\pm} = \frac{m_\pm}{M} \v P \pm\! \sqrt{2 m E} \frac{\v x}{\left|\v x\right|} \qquad , \qquad
	E_{\pm} = \frac{m_{\pm}}{M} E_{\text{CM}} + \frac{m_{\mp}}{M} E \pm \sqrt{\frac{4 m}{M} E E_{\text{CM}}} \cos\theta_{Px}
	\end{equation}
	
	The screening potential $U$ is then written as in \cref{eqUExpression}, using \cref{eqPrimaryReactionsReactantsEnergyGain,eqPrimaryReactionsCoMEnergyGain}.
	Its average over $\cos\theta_{P0}$ and $\cos\theta_{x0}$ is straightforward, since $U$ depends on those linearly through $\Delta E_{\pm}$.
	For simplicity, consider only the case where $Z_1 = Z_2 = Z$ and $m_1 = m_2$. Then $\Delta E_{\pm}$ does not depend on $\theta_{Px}$, so:
	\begin{equation}
	\frac{1}{4} \int \d\cos\theta_{P0} \d\cos\theta_{x0} \, \Delta E_{\pm} = \frac{1}{2} \average{\Delta E}_{|r_0}
	\end{equation}
	where $\average{\Delta E}_{|r_0} = \average{\Delta E_+}_{|r_0} + \average{\Delta E_-}_{|r_0}$.
	The average center-of-mass energy gain is:
	\begin{equation}
	\average{\Delta E_{\text{CM}}}_{|r_0} \approx
	\int_{-1}^{1} \frac{\d\cos\theta_{Px}}{2} \left[ \frac{E_{\text{CM}}^2 - E^2}{4 E_+ E_-} + 1 \right] \frac{\average{\Delta E}_{|r_0}}{2}
	\end{equation}
	so the average screening potential is:
	\begin{equation}
	\average{U}_{|r_0} \approx \frac{k_e q_e^2}{2} \tilde\rho_s Z x^2 (\gamma-1)
	\left[ 1 + \frac{ 1 - \gamma }{ 2 \sqrt{\gamma} } \ln\left(\frac{ 1 + \sqrt{\gamma} }{ \left|1 - \sqrt{\gamma}\right| }\right) \right]
	\end{equation}
	Further averaging over $x$ as in \cref{eqExplicitxAverage} gives (the mean free path $\lambda$ is the same for both reactants, since they are equal):
	\begin{equation}\label{eqScreenPotArbitraryMotionxAveraged}
	\average{U}_{|r_0, \, \lambda, \, \gamma, \, m_+=m_-, \, Z_+=Z_- } \approx
	\left\lbrace \begin{aligned}
	& k_e q_e^2 \tilde\rho_s Z \lambda^2 \frac{\gamma(\gamma-1)}{ 4 \left(\gamma+\frac{1}{3}\right)^2 } \left[ 1 + \frac{ 1 - \gamma }{ 2 \sqrt{\gamma} } \ln\left(\frac{ 1 + \sqrt{\gamma} }{ \left|1 - \sqrt{\gamma}\right| }\right) \right] &\text{if } & \gamma \geq 1 \\
	& k_e q_e^2 \tilde\rho_s Z \lambda^2 \frac{\gamma-1}{ 4 \left(1+\frac{\gamma}{3}\right)^2 }  \left[ 1 + \frac{ 1 - \gamma }{ 2 \sqrt{\gamma} } \ln\left(\frac{ 1 + \sqrt{\gamma} }{ \left|1 - \sqrt{\gamma}\right| }\right) \right] &\text{if } & \gamma \leq 1
	\end{aligned} \right.
	\end{equation}
	In order to obtain an energy-independent screening potential, this will be also averaged over $E$ and $E_{\text{CM}}$ distributions, $\Phi_E$ and $\Phi_{E_{\text{CM}}}$, which by assumption are Maxwell-Boltzmann distributions with equal temperature.
	Since \cref{eqScreenPotArbitraryMotionxAveraged} depends on reactants energies only through $\gamma$, it is:
	\begin{equation}
	\average{U}_{|r_0, \, \lambda}
	= \int_{0}^{+\infty} \d E \int_0^{+\infty} \d E_{CM} \Phi(E) \Phi(E_{CM}) \average{U}_{|r_0, \, \lambda, \, \gamma}
	= \frac{8}{\pi} \int_0^{+\infty} \d \gamma \frac{\sqrt{\gamma}}{\left(1+\gamma\right)^3} \average{U}_{|r_0, \, \lambda, \, \gamma}
	\end{equation}
	which can be evaluated numerically, giving:
	\begin{equation}\label{eqScreenPotArbitraryMotionEnergyAveraged}
	\average{U}_{|r_0, \, \lambda, \, m_+=m_- ,\, Z_+=Z_-=Z } \approx -\num{0.176} k_e q_e^2 Z \tilde\rho_s \frac{\lambda^2}{2}
	\end{equation}
	
	Let $Z_s q_e$ ($q_e$ is the proton charge) be the average net charge per nucleus in the sphere, which is of course not granted or actually expected to be independent of the plasma conditions.
	The quantity $\tilde\rho_s$, defined in \cref{sezModelICF}, is then:
	\begin{equation}\label{eqDefinitionZs}
	\tilde\rho_s = \frac{4}{3} \pi Z_s n_i
	\end{equation}
	where $n_i$ is the sphere total number density of nuclei.
	Qualitatively, $Z_s$ is expected to be very small, of the order of \num{e-4} or less (see \cite[sec.\ 2.4.A]{PerrottaTesiSSC}).

\subsubsection{Inertial confinement fusion plasma}

\begin{figure}[tbp]
		\centering%
		\includegraphics[keepaspectratio = true, width=\linewidth, height=0.4\textheight]{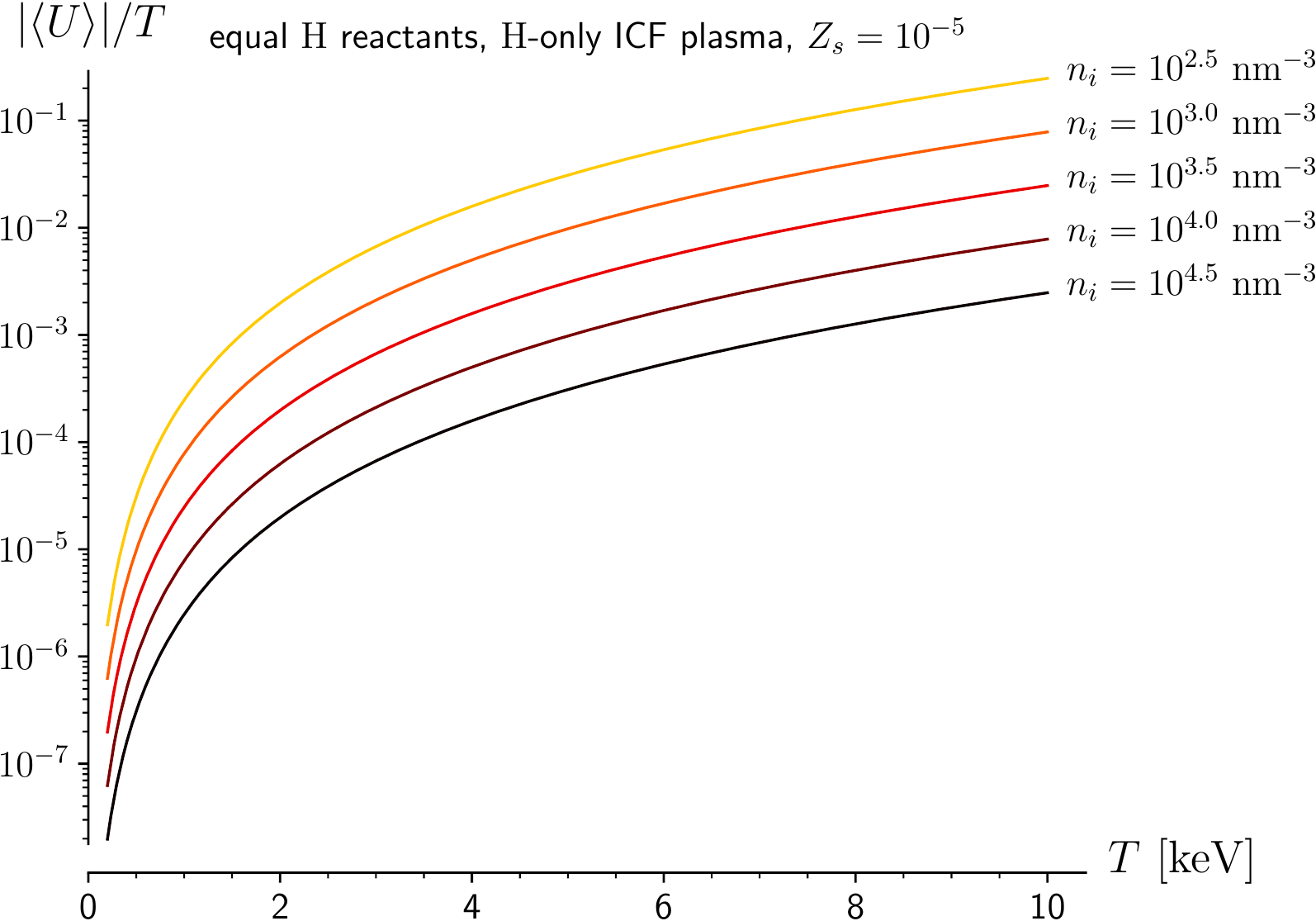}%
		\caption[Screening potential for equal hydrogen reactants]%
		{\label{fig_PotScreenHinH}%
			Absolute value of the screening potential in \cref{eqScreenPotArbitraryMotionEnergyAveragedICF} divided by the hotspot temperature $T$,
			as a function of $T$, for an ICF plasma made of only hydrogen isotopes, with
			a net charge per nucleus of $\num{e-5}$ proton charges (see \scref{eqDefinitionZs}). Each curve represents a different number density of nuclei $n_i$ as marked by the labels.}
	\end{figure}

\begin{figure}[tbp]
		\centering%
		\includegraphics[keepaspectratio = true, width=\linewidth, height=0.4\textheight]{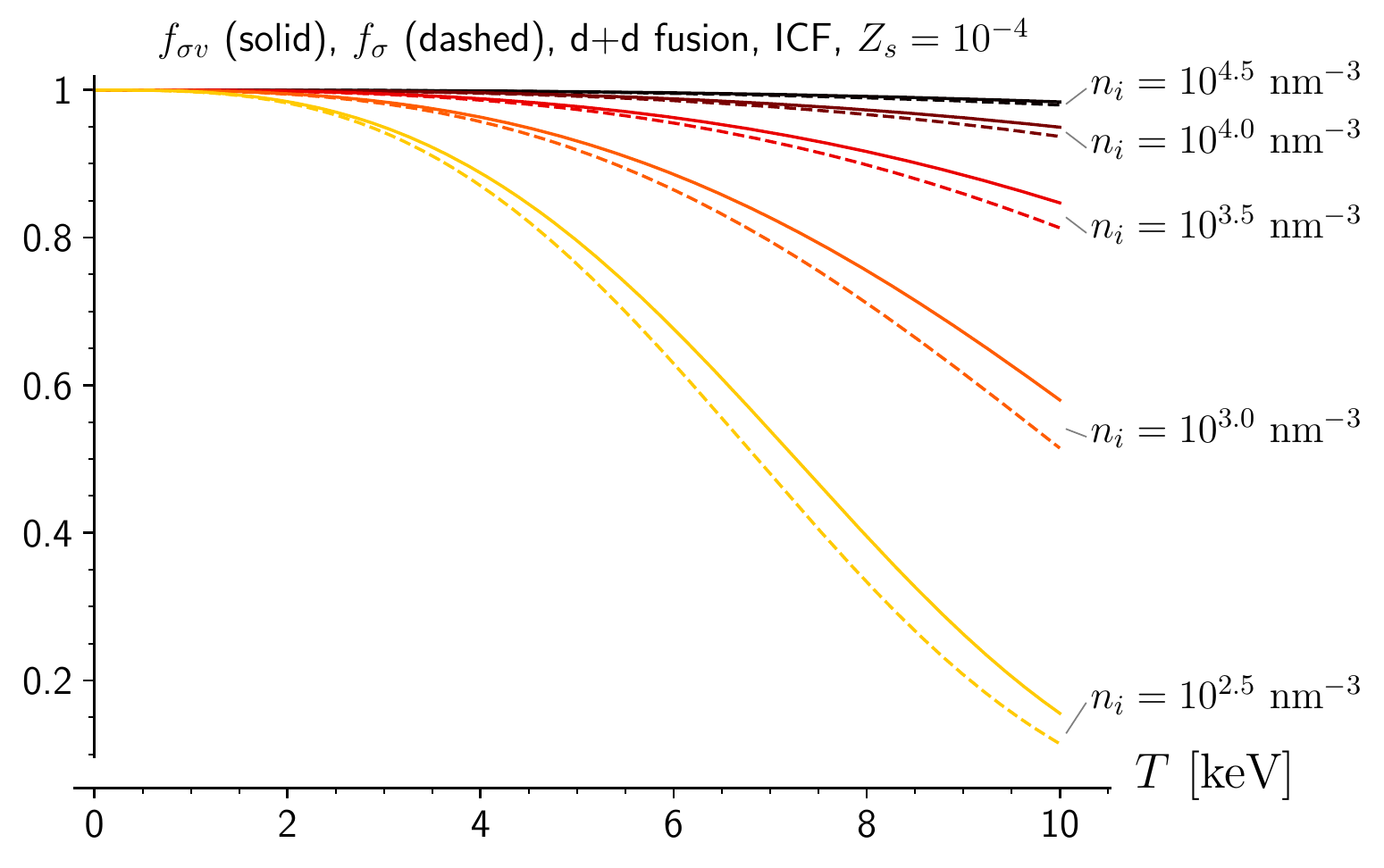}%
		\vfill
		\includegraphics[keepaspectratio = true, width=\linewidth, height=0.4\textheight]{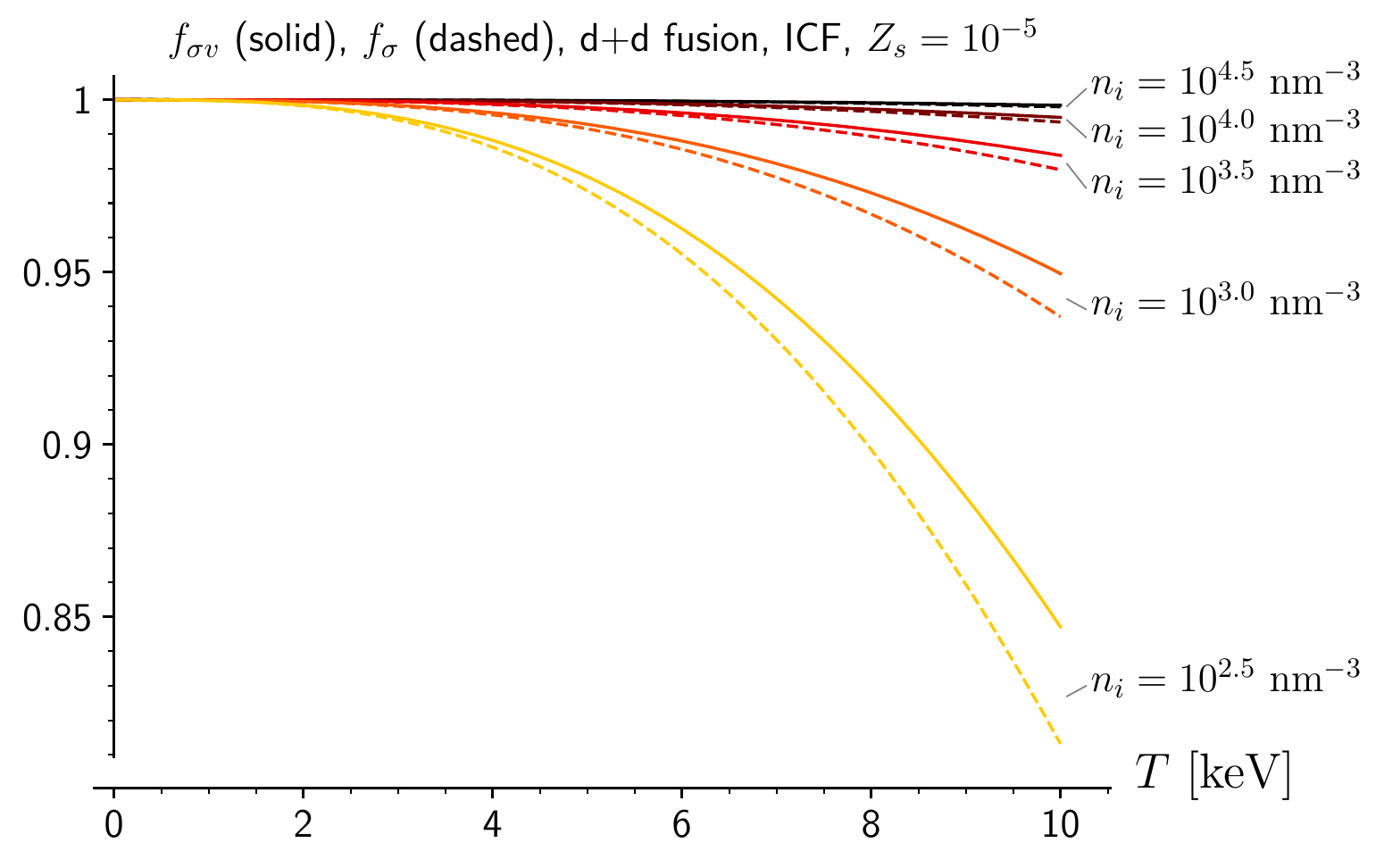}%
		\caption[Rate and cross-section enhancement factors for \texorpdfstring{$\nuclide{d}+\nuclide{d}$}{d+d}]%
		{\label{fig_Fusionrateenhancementfactorsdd}%
			Enhancement factors in \cref{eqScreenedRateSaddle} for the rate per particle pair, $f_{\sigma v}$ (solid lines), and the average cross-section, $f_{\sigma}$ (dashed lines),
			calculated using \cref{eqScreenPotArbitraryMotionEnergyAveragedICF}, as a function of the hotspot temperature $T$, for $\nuclide{d}+\nuclide{d}$ fusion in an ICF plasma made of only hydrogen isotopes. Each color represents a different number density of nuclei $n_i$ as marked by the labels. Top panel refers to a net charge number per nucleus of $Z_s = \num{e-4}$ (see \scref{eqDefinitionZs}) and bottom panel to $Z_s = \num{e-5}$.}%
	\end{figure}

\begin{figure}[tbp]
		\centering%
		\includegraphics[keepaspectratio = true, width=\linewidth, height=0.4\textheight]{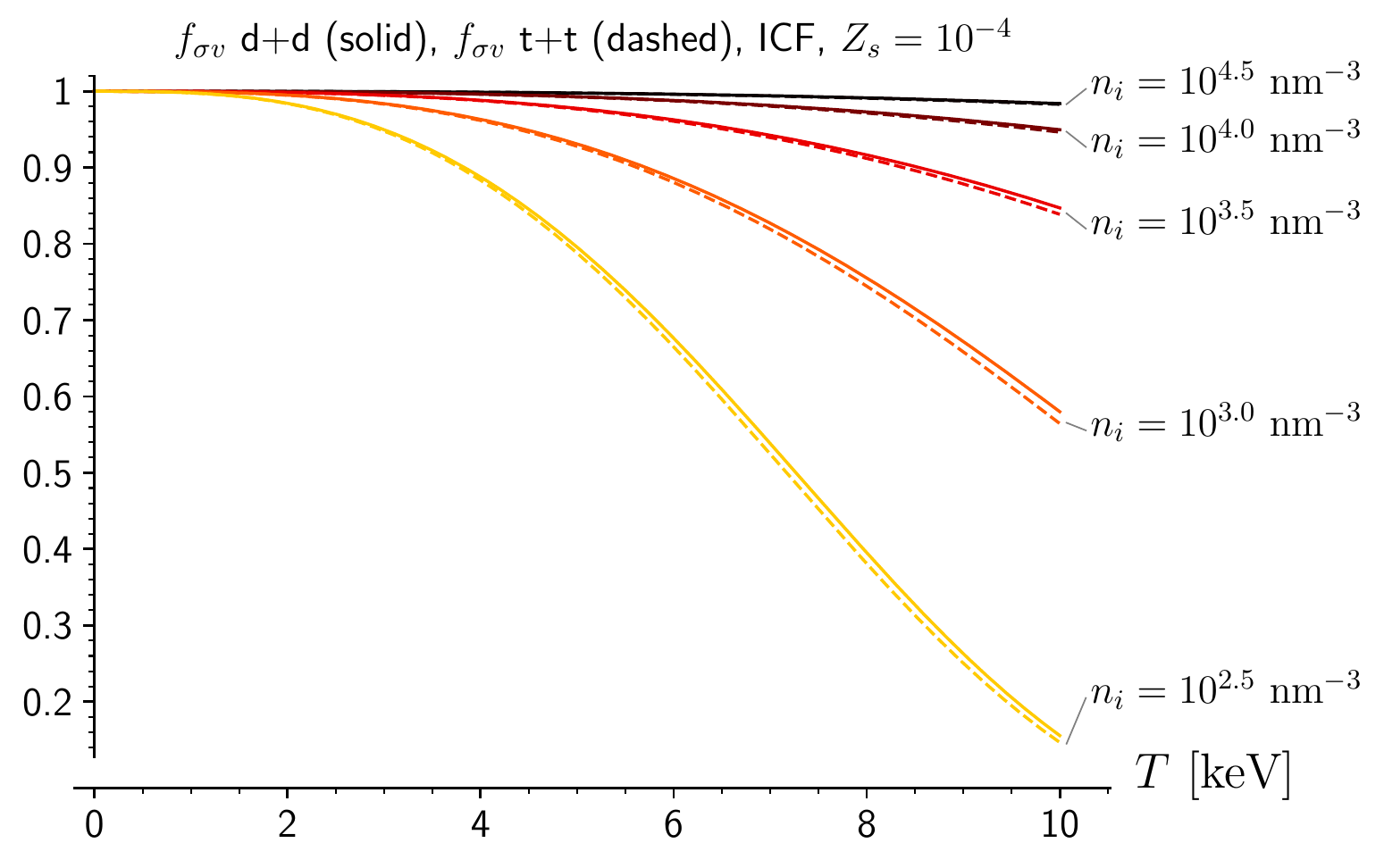}%
		\caption[Rate enhancement factors for \texorpdfstring{$\nuclide{d}+\nuclide{d}$}{d+d} and \texorpdfstring{$\nuclide{t}+\nuclide{t}$}{t+t}]%
		{\label{fig_Fusionrateenhancementfactorsddtt}%
			Rate enhancement factor $f_{\sigma v}$ in \cref{eqScreenedRateSaddle},
			calculated using \cref{eqScreenPotArbitraryMotionEnergyAveragedICF},
			as a function of the hotspot temperature $T$, for $\nuclide{d}+\nuclide{d}$ (solid lines) and $\nuclide{t}+\nuclide{t}$ (dashed lines) fusion in an ICF plasma
			made of only hydrogen isotopes, with a net charge per nucleus of $\num{e-4}$ proton charges (see \scref{eqDefinitionZs}).
			Each color represents a different number density of nuclei $n_i$ as marked by the labels.}
	\end{figure}

	Using \cref{eqDefinitionZs} and the expression in \cref{eqMeanFreePathICF} for the mean free path in an ICF hotspot, \cref{eqScreenPotArbitraryMotionEnergyAveraged} becomes:
	\begin{equation}\label{eqScreenPotArbitraryMotionEnergyAveragedICF}
	\average{U}_{|r_0, \, \lambda_{\text{ICF}} ,\, m_+=m_- ,\, Z_+=Z_-=Z } \approx
	-\frac{\num{0.176}}{6 \pi \ln^2(\Lambda)} \frac{T^4}{Z^3 (k_e q_e^2)^3} \frac{n_i}{\left(\sum_a Z_a^2 n_a\right)^2} Z_s
	\end{equation}
	\Cref{fig_PotScreenHinH} shows the ratio of the $U$ in \cref{eqScreenPotArbitraryMotionEnergyAveragedICF} to the hotspot temperature, for a pair of equal hydrogen reactants in an ICF hotspot made only of hydrogen isotopes and electrons.
	In \cref{fig_Fusionrateenhancementfactorsdd} there are plots of the corresponding reaction rate and average cross-section enhancement factors from \cref{eqScreenedRateSaddle}, $f_{\sigma v}$ and $f_{\sigma}$, for $\nuclide{d}+\nuclide{d}$ fusion.
	The results are quite similar for $f_{\sigma v}$ and $f_{\sigma}$, as for both of them the dominant contribution is given by the exponential term in \scref{eqScreenedRateSaddle}, to which the differing $U/E_0$ part is only a second-order correction.
	For the same reason, the enhancement factors change very little for other pairs of identical hydrogen isotopes, since reactants mass influences only $E_0$ in the equations:
	\cref{fig_Fusionrateenhancementfactorsddtt} compares the rate enhancement factor for $\nuclide{d}+\nuclide{d}$ and $\nuclide{t}+\nuclide{t}$ fusion under otherwise identical conditions.
	Note also that essentially only the temperature and the ratio $Z_s/n_i$ are relevant for $U$ and $f$,
	thus, for any $C \ll 1/Z_s$,
	it is:
	\begin{equation}\label{eqRelationEnhancementDifferentZsni}
	f(T, n_i,Z_s) \approx f(T, C n_i, C Z_s)
	\end{equation}

\subsubsection{Coulomb explosion plasma}
	
\begin{figure}[tbp]
		\centering%
		\includegraphics[keepaspectratio = true, width=\linewidth, height=0.4\textheight]{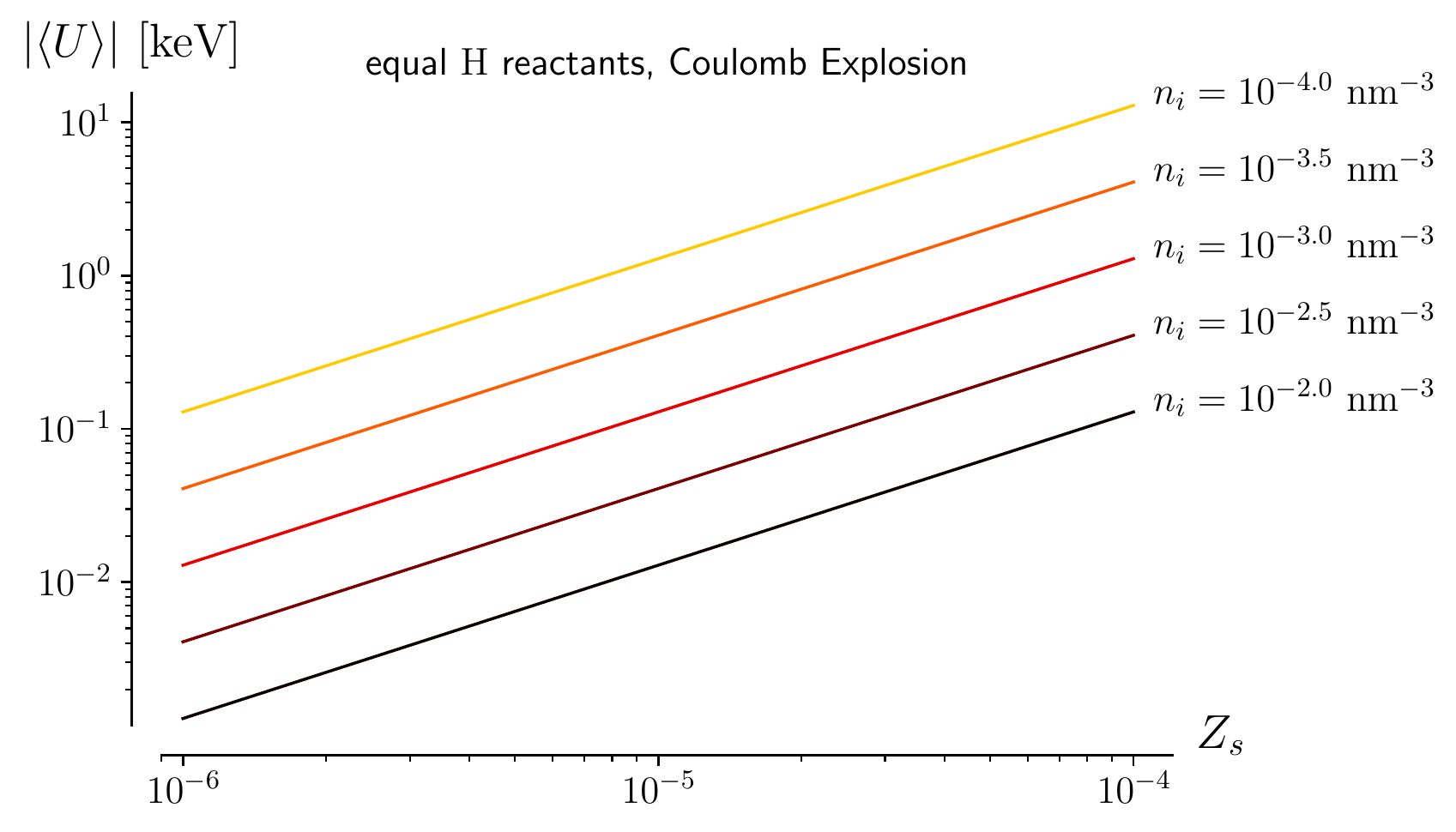}%
		\caption{\label{fig_PotScreeningCE}%
			Absolute value of the screening potential in \cref{eqPotScreeningCE} as a function of the net charge number per nucleus $Z_s$ (see \scref{eqDefinitionZs}), for a pair of equal hydrogen reactants in a Coulomb explosion plasma. Each curve represents a different number density of nuclei $n_i$ as marked by the labels.}
	\end{figure}

\begin{figure}[tbp]
		\centering%
		\includegraphics[keepaspectratio = true, width=\linewidth, height=0.4\textheight]{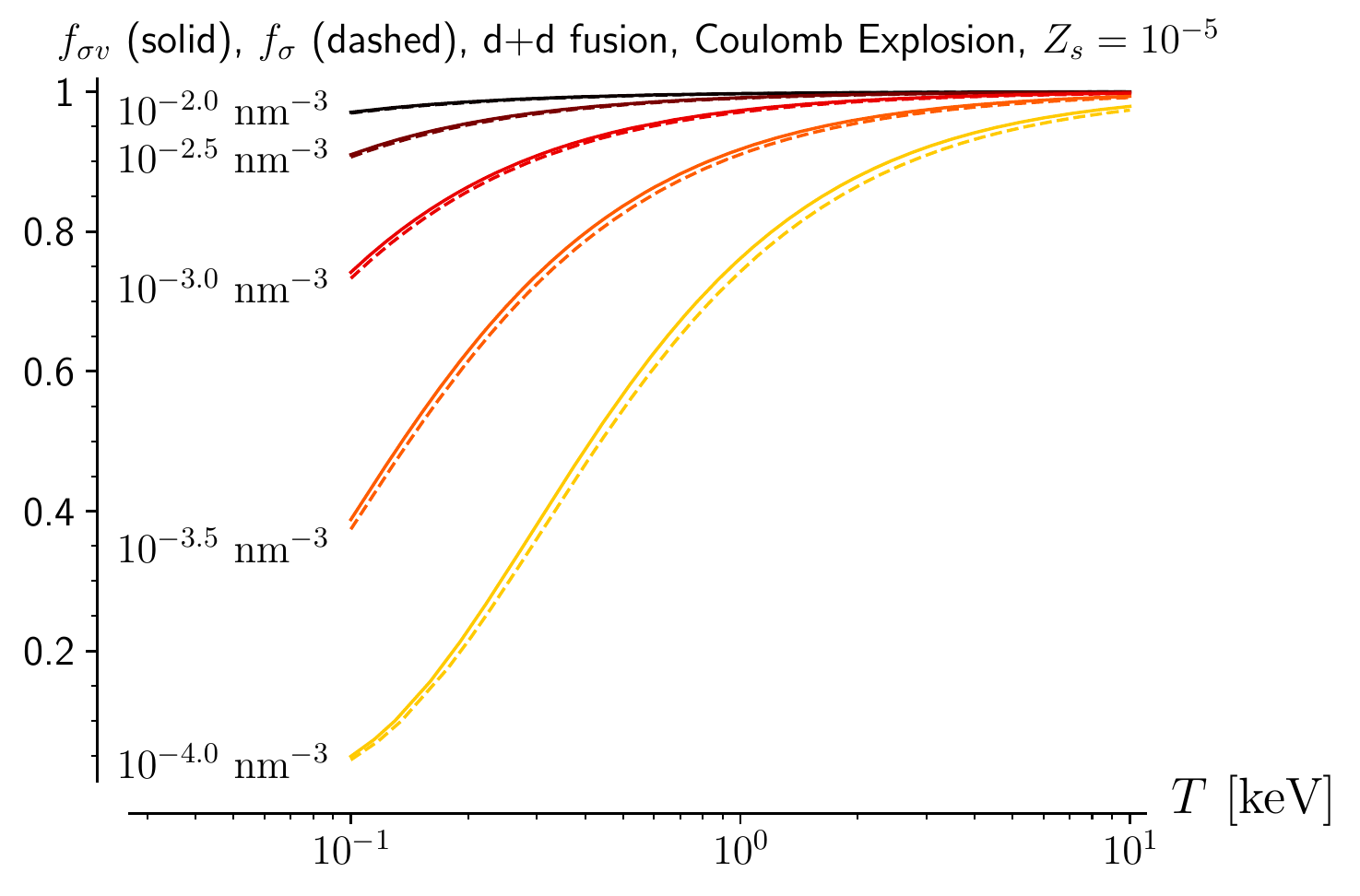}%
		\caption{\label{fig_EnhancementCE}%
			Enhancement factors in \cref{eqScreenedRateSaddle} for the rate per particle pair, $f_{\sigma v}$ (solid lines), and the average cross-section, $f_{\sigma}$ (dashed lines),
			calculated using \cref{eqPotScreeningCE}, as a function of the reactants temperature $T$ (in logarithmic scale), for $\nuclide{d}+\nuclide{d}$ fusion in a Coulomb explosion plasma,
			with a net charge per nucleus of $\num{e-5}$ proton charges (see \scref{eqDefinitionZs}). Each color represents a different number density of nuclei $n_i$ as marked by the labels.}
	\end{figure}

	\emph{Beam-beam} reactions in CE systems are collisions between accelerated ions following a Maxwell-Boltzmann energy distribution at given $T$, i.e.\ the equivalent of ICF primary reactions.
	To describe those, \cref{eqMeanFreePathCE} is inserted in \cref{eqScreenPotArbitraryMotionEnergyAveraged} (the average screening potential for a pair of identical reactants with Maxwell-Boltzmann energy distribution):
	\begin{equation}\label{eqPotScreeningCE}
	\average{U}_{|r_0, \, \lambda_{\text{CE}} ,\, m_+=m_- ,\, Z_+=Z_-=Z} \approx -\num{0.176} k_e q_e^2 Z \frac{2 N_{ic}^{2/3}}{3 \pi r_s^4} \frac{Z_s}{n_i}
	\end{equation}
	\Cref{fig_PotScreeningCE,fig_EnhancementCE} show, for this scenario, plots analogous to those in \cref{fig_PotScreenHinH,fig_Fusionrateenhancementfactorsdd}, again for any pair of equal hydrogen reactants and for $\nuclide{d}+\nuclide{d}$ fusion respectively (the plasma composition is here irrelevant).
	As in the inertial confinement fusion case, the difference between the rate and average cross-section enhancement factors, $f_{\sigma v}$ and $f_{\sigma}$, as well as the difference between the results for different reactants mass, are rather small, for the same reasons. \Cref{eqRelationEnhancementDifferentZsni} is here manifestly true for any real $C$.
	Comparing the ICF and the CE case, note the qualitatively opposite behavior of the enhancement factors
	with respect to the temperature, due to the fact that $\lambda_{\text{CE}}$ and $\average{U}_{\text{CE}}$ are
	independent of $T$,
	while $\lambda_{\text{ICF}} \propto T^2$ (until \cref{eqMeanFreePathICF} breaks down for too high $\lambda$).
	
\section{Summary and conclusions}\label{sezSummary}

	This work investigated how the cross-section of nuclear fusion reactions at ultra-low energy ($\lesssim \SI{10}{\keV}$) can be modified in a positively charged environment.
	The direction and importance of those variations critically depends on the reactants and plasma conditions, as can be seen by comparing \cref{eqAverageScreeningPotReactantRest,eqScreenPotArbitraryMotionEnergyAveragedICF,eqPotScreeningCE}.
	In particular, it was found that primary (and beam-beam) reactions are hindered by the analyzed environment, while secondary (and beam-target) fusions are moderately enhanced.
	Rate and average cross-section enhancement factors (ratios of screened to bare quantities) were explicitly computed for primary reactions between identical reactants.
	Remarkably distinct results were in this way obtained for inertial confinement fusion (ICF) and Coulomb explosion (CE) systems, due to the very different plasma structure in the two cases.
	Regarding secondary reactions, to quantitatively predict the effects magnitude a more precise numerical evaluation is needed, which may be included in future studies.
	
	Clearly, the modifications are more substantial for greater plasma net charge densities (specifically, results were expressed in terms of the average net charge per nucleus, $Z_s q_e$, see \cref{eqDefinitionZs}).
	It is however emphasized that important variations were found, for some plausible nucleus number density and temperature values, already at small charge densities ($Z_s q_e = \num{e-5}$ proton charges).
	The ions mean free path against significant deflections, $\lambda$, (see \cref{eqMeanFreePathICF,eqMeanFreePathCE}) also plays a prominent role.
	For fixed $Z_s$, the present model predicts less considerable alterations for denser plasmas, as in those $\lambda$ is reduced in both the ICF and CE scenarios.
	In CE plasmas, ‘hotter' reactants produce smaller modifications.
	For ICF systems, the opposite behavior with respect to the temperature was found, because more energetic particles can travel greater distances in the hotspot before enduring a given deflection.
	
	The $\lambda^2$ factor in the average screening potential (see \cref{eqAverageScreenPotRestCM,eqScreenPotArbitraryMotionEnergyAveraged}) is responsible for most of the “non-trivial” features of the enhancement factors. It is remarked that said factor is ultimately produced by the behavior of the uniform sphere electrostatic potential, $V(r)$, in \cref{eqElectrostaticPotentialUniformSphere}.
	The $r^2$ dependence ($r$ is the distance from the sphere center) in the potential causes all energy gains to have a term proportional to the initial reactants distance norm, $4 x^2 = \left|\v r_- - \v r_+\right|^2$.
	Its sign and magnitude (possibly zero, as in \cref{eqScreeningPotReactantRest}) depend on several parameters, but this is always the only term left once the average screening potential is computed.
	In turn, the $x^2$ dependence gives rise to the sought $\lambda^2$, through the integration over the distance covered on average by a reactant (linear in $x$, see \cref{eqAngleAverageReactantCoveredLenght}).
	
	An important subsequent step to this study will be to compare the predictions of the present model with existing experimental data and simulations.
	Doing so will determine whether reaction rates experimentally extracted without taking into account any cross-section modification do follow the trends suggested in this paper.
	Furthermore, it will allow to understand if, and to what extent, the hindrance effects here discussed can reduce the discrepancies often appearing between simulated and measured fusion yields (see e.g. \cite{Gopalaswamy2019}).
	The difficulty in performing these comparisons lies in the fact that not all physical quantities required by the present model are actually measured or simulated.
	A careful evaluation of each data set will therefore be necessary, to assess whether a charged environment can be expected in a given setup, and if uncontrolled parameters (as the net charge and the confinement time) appear reasonably consistent between different compared results.
\section*{Acknowledgements}

	This work was supported by the US National Nuclear Security Administration under Grant No.\ {de-na0003841} (CENTAUR).
	
	One of us (S.~S.~P.) thanks the Scuola Superiore di Catania and the Università degli studi di Catania for financial support and the Cyclotron institute, Texas A\&M University for the warm hospitality and local support during his stay while this project was developed.
\bibliography{bibliography}%
\end{document}